\documentclass[aps,prd,twocolumn,superscriptaddress,showkeys,showpacs]{revtex4}
\usepackage[utf8]{inputenc}
\usepackage[T1]{fontenc}
\usepackage{pdfsync}
\usepackage{mathptmx}
\usepackage{epsfig,epsf}
\usepackage{amsmath}
\usepackage{amsthm}
\usepackage{amsfonts}
\usepackage{amssymb}
\usepackage{dsfont}

\usepackage{slashed}

\usepackage[active]{srcltx}

\newcommand{\be}{\begin{equation}}
\newcommand{\ee}{\end{equation}}
\newcommand{\ba}{\begin{eqnarray}}
\newcommand{\ea}{\end{eqnarray}}
\newcommand{\baa}{\begin{eqnarray*}}
\newcommand{\btab}{\begin{tabular}}
\newcommand{\etab}{\end{tabular}}
\newcommand{\eaa}{\end{eqnarray*}}

\newcommand \vev [1] {\langle{#1}\rangle}
\newcommand \VEV [1] {\left\langle{#1}\right\rangle}

\def\inbar{\,\vrule height1.5ex width.4pt depth0pt}
\def\IC{\relax\hbox{$\inbar\kern-.3em{\rm C}$}}
\def\IZ{\relax{\hbox{\cmss Z\kern-.4em Z}}}
\def\IR{{\hbox{{\rm I}\kern-.2em\hbox{\rm R}}}}

\def\IP{{\hbox{{\rm I}\kern-.2em\hbox{\rm P}}}}
\def\II{\hbox{{1}\kern-.25em\hbox{l}}}

\newcommand \widebar [1] {\overline{#1}}

\begin{document}

\title{Higher twist nucleon distribution amplitudes in Wandzura-Wilczek approximation}


\date{\today}

\author{ I.~V.~Anikin}
\affiliation{Institut f\"ur Theoretische Physik, Universit\"at
   Regensburg,D-93040 Regensburg, Germany}
\affiliation{Bogoliubov Laboratory of Theoretical Physics, JINR, 141980 Dubna, Russia}
\author{ A.~N.~Manashov}
\affiliation{Institut f\"ur Theoretische Physik, Universit\"at
   Regensburg,D-93040 Regensburg, Germany}
\affiliation{Department of Theoretical Physics,  St.-Petersburg State
University,
199034, St.-Petersburg, Russia}

\date{\today}

\begin{abstract}
We derive the higher twist (four and five) nucleon distribution amplitudes
in the Wandzura-Wilczek approximation.
Our method is based on an analysis of the conformal expansion of  nonlocal operators
in the spinor formalism.
\end{abstract}


\maketitle


%
\section{Introduction}

Hard exclusive processes provide  an opportunity to  access  the internal structure of
hadrons. As usual the processes with nucleons are of special importance due to their experimental
availability. The theoretical description of exclusive processes is based on the QCD
factorization
approach~\cite{Chernyak:1977as,Chernyak:1977fk,Efremov:1979qk,Efremov:1978rn,Lepage:1979zb}.
It introduces  a notion of the hadron distribution amplitudes (DAs)
which can be thought of as momentum
fraction distributions  of partons in configurations with a fixed
number of Fock constituents.
Hadron DAs are customary defined as hadron matrix elements of the corresponding nonlocal
operators which can be classified according to their twist.
In the  factorization approach the dominant contribution to an amplitude of exclusive process
in the large  $Q^2$ limit comes from DAs of lowest possible twist -- minimal number of constituents
-- two for mesons and
three for baryons. The QCD approach, however, faces conceptual
difficulties in application to baryons (see
Refs.~\cite{Duncan:1979hi,Duncan:1979ny,Milshtein:1981cy,Milshtein:1982js,Kivel:2010ns})
and fails to provide a quantitative description for $Q^2$ accessible in the current and
planned experiments. In particular, this concerns the description of the electromagnetic
nucleon form factors.

Attempts to get a more realistic description of the nucleon electromagnetic form factors
within the light-cone sum rule (LCSR)
approach~\cite{Balitsky:1986st,Balitsky:1989ry,Chernyak:1990ag} by taking into account
power suppressed corrections were undertaken in
Refs.~\cite{Braun:2001tj,Braun:2006hz,Lenz:2003tq,Aliev:2008cs} at the leading order in
$\alpha_s$ and in~Refs.~\cite{PassekKumericki:2008sj,Anikin:2013aka} at the
next-to-leading order. In this approach the higher twist nucleon DAs enter the expressions
for the form factors as nonperturbative inputs.
However, the present day
knowledge of the nucleon DAs is quite poor. Only the leading twist (twist$-3$) nucleon DA
is known with some degree of certainty (see e.g. Ref.~\cite{Anikin:2013aka} and reference
therein). At the same time, it is well known that the higher twist DAs can be split into
two parts --
 the kinematical and dynamical ones. The latter defines a genuinely new nonperturbative
function while the former can be expressed in terms of the DAs of lower twist.
The approximation in which only the kinematical part of a DA is taken into account
is usually referred to as the Wandzura-Wilczek (WW)
approximation.

In the case of meson DAs, a procedure for reconstructing of the Wandzura-Wilczek contribution
is well known (see e.g. Ref.~\cite{Ball:1998ff}). The WW
contributions to the nonforward matrix elements of twist three operators were obtained
in~\cite{Belitsky:2000vx,Radyushkin:2000jy,Kivel:2000rb,Anikin:2001ge,Geyer:2004bx}.
Unfortunately, these methods are not applicable
to the nucleon DAs that are determined by matrix elements of three-quark operators.
As a consequence, the WW relations for the nucleon DAs were known only for
few first moments~\cite{Braun:2001tj,Braun:2006hz}.

In the present paper we develop a technique for calculation of the WW corrections to higher twist DAs.
Our analysis relies on the spinor formalism which proves to be very effective for studies of the
nucleon DAs, see Ref.~\cite{Braun:2000kw}.
In combination with the conformal wave expansion for nonlocal operators
it allows us to simplify substantially the derivation of the twist$-4$ WW terms
given in~\cite{Braun:2000kw} and calculate the WW corrections to the twist$-5$ DAs.
Making use of a new summation formula for conformal series we avoid calculation of certain
normalization coefficients that requires knowledge of the evolution Hamiltonian for the
corresponding operators.

The paper is organized as follows: Sect.~\ref{sect:DAs} is introductory.
We fix the notations and provide  definitions for various nucleon DAs in the spinor formalism.
In sect.~\ref{PF4} we explain our approach on the example of the WW contribution to the twist$-4$
DAs. Sect.~\ref{PF5} contains details of the calculation of the twist$-5$ DAs.
The results for all nucleon DAs up to twist five are collected in
sect.~\ref{Results}. In several Appendices we explain some technical issues.

\section{Nucleon distribution amplitudes}\label{sect:DAs}

The three-quark nucleon DAs of different twist are defined
by matrix elements of the corresponding three quark operators, see Refs.~\cite{Braun:2000kw,Braun:2001tj}.
Below we present definitions for the relevant DAs in the spinor formalism.
We will follow closely the notations of Ref.~\cite{Braun:2011aw}.

The leading twist DA is defined as
\begin{multline}\label{DA3}
\vev{0|\epsilon^{ijk}u^{\downarrow i}_+(z_1n)u^{\uparrow j}_+(z_2n) d^{\downarrow
k}_+(z_3n)|P}=
 \\
 =
-\frac12 (p n)\, N^{\downarrow}_+\,\int \mathcal{D}x \, e^{-i(pn)\,\sum
{x_iz_i}}\,\Phi_3(x)\,.
\end{multline}
The integration in~(\ref{DA3}) goes over the simplex, 
i.e.
\begin{align}
\mathcal{D}x=dx_1\,dx_2\,dx_3\,\delta(1-x_1-x_2-x_3).
\end{align}
The up and down  spinors, $q^{\uparrow(\downarrow)}$,
are the two component Weyl spinors,
\begin{align}\label{}
q=
\begin{pmatrix}q^{\downarrow}\\
q^{\uparrow}
\end{pmatrix}\,, &&q^{\uparrow(\downarrow)}=\frac12(1\pm\gamma_5) q\,,
\end{align}
and similarly for the nucleon spinor $N$.
The vector $n$ is an auxiliary light-like vector ($n^2=0$). It can be parameterized  by a  Weyl spinor
$\lambda$ as follows
\begin{align}
n_\mu\sigma^\mu_{\alpha\dot\alpha}=\lambda_\alpha\bar\lambda_{\dot\alpha},
\end{align}
where $\bar\lambda=\lambda^\dagger$.
For later convenience we introduce the second light-like vector $\tilde n$, such that $(n\tilde n)\neq 0$
and denote the corresponding auxiliary spinor by $\mu$,
($\tilde n_\mu\sigma^\mu_{\alpha\dot\alpha}=\mu_\alpha\bar\mu_{\dot\alpha}$).
We introduce shorthand notations
for  projections of the quark fields onto the auxiliary spinors $\lambda$ and $\mu$
\begin{align*}
q^{\downarrow}_+=\lambda^\alpha q^{\downarrow}_\alpha,&&
q^{\downarrow}_-=\mu^\alpha q^{\downarrow}_\alpha, &&
q^{\uparrow}_+=\bar\lambda^{\dot\alpha} q^{\uparrow}_{\dot\alpha}, &&
q^{\uparrow}_-=\bar\mu^{\dot\alpha} q^{\uparrow}_{\dot\alpha}.
\end{align*}
We accept the conventions of Refs.~(\cite{Braun:2008ia,Braun:2009vc})
for rising and lowering spinor indices. 

The nucleon DAs of twist$-4$, $\Phi_4,\,\Psi_4,\,\Xi_4$, and
of twist$-5$, $\Phi_5,\,\Psi_5,\,\Xi_5$ were defined in
Refs.~\cite{Braun:2000kw,Braun:2001tj}.
Below we present definitions for these DAs in the spinor formalism that is more convenient
for
further analysis.
We get
\begin{align}\label{DA4}
&\vev{0|\epsilon^{ijk} u_+^{\downarrow i}(z_1) u_+^{\uparrow j}(z_2) d_{-}^{\downarrow k}(z_3)|P}=
\notag\\
&\phantom{dilogarithm}=
\frac{1}4(\mu\lambda) m_N N^{\uparrow}_+
\int Dx \,e^{-i(pn)\sum z_k x_k}\,\Phi_4(x)\,,
\notag\\[2mm]
&\vev{0|\epsilon^{ijk} u_+^{\uparrow i}(z_1) u_-^{\downarrow j}(z_2) d_{+}^{\downarrow k}(z_3)|P}=
\notag\\
&\phantom{dilogarithm}=
\frac{1}4(\mu\lambda) m_N N^{\uparrow}_+
\int Dx \,e^{-i(pn)\sum z_k x_k}\,\Psi_4(x)\,,
\notag\\[2mm]
&\vev{0|\epsilon^{ijk} u_-^{\downarrow i}(z_1) u_+^{\downarrow j}(z_2) d_{+}^{\downarrow k}(z_3)|P}=
\notag\\
&\phantom{dilogarithm}=
\frac{1}4(\mu\lambda) m_N N^{\downarrow}_+
\int Dx \,e^{-i(pn)\sum z_k x_k}\,\Xi_4(x)\,
\end{align}
for the twist$-4$ DAs and
\begin{align}\label{DA5}
&\vev{0|\epsilon^{ijk} u_-^{\downarrow i}(z_1) u_-^{\uparrow j}(z_2) d_{+}^{\downarrow k}(z_3)|P}=
\notag\\
&\phantom{dilog}=
-\frac{1}4(\mu\lambda) m_N N^{\uparrow}_-(P)
\int Dx \,e^{-i(pn)\sum z_k x_k}\,\Phi_5(x)\,,
\notag\\[2mm]
&\vev{0|\epsilon^{ijk} u_-^{\uparrow i}(z_1) u_+^{\downarrow j}(z_2) d_{-}^{\downarrow k}(z_3)|P}=
\notag\\
&\phantom{diloga}=
-\frac{1}4(\mu\lambda) m_N N^{\uparrow}_-(P)
\int Dx \,e^{-i(pn)\sum z_k x_k}\,\Psi_5(x)\,,
\notag\\[2mm]
&\vev{0|\epsilon^{ijk} u_+^{\downarrow i}(z_1) u_-^{\downarrow j}(z_2) d_{-}^{\downarrow k}(z_3)|P}=
\notag\\
&
\phantom{dilog}=-\frac{1}4(\mu\lambda) m_N N^{\downarrow}_-(P)
\int Dx \,e^{-i(pn)\sum z_k x_k}\,\Xi_5(x)\,.
\end{align}
for the twist$-5$ DAs.
Here $m_N$ stands for the nucleon mass and $q_\pm(z)\equiv q_\pm(zn)$.
Notice that these definitions differ by a sign from those given in Ref.~\cite{Braun:2008ia}
due to the nonstandard charge conjugation matrix $C$ used there.

The twist$-3$ DA $\Phi_3$ can be represented as a series
\begin{align}\label{Phi3}
\Phi_3(x)=x_1x_2x_3\sum_{N,q} c_{Nq}\,\phi_{Nq}(\mu_R) \,P_{Nq}(x_1,x_2,x_3)\,,
\end{align}
where $\mu_R$ is the renormalization scale. The common prefactor  is dictated by the conformal symmetry.
$P_{Nq}$ are the homogeneous polynomials of degree $N$,
$P_{Nq}(s x)=s^N P_{Nq}(x)$,
that form an orthogonal system
\begin{align}\label{}
\int \mathcal{D}x\, x_1 x_2 x_3\,P_{Nq}(x)P^\dagger_{Nq'}(x)=\delta_{qq'}c_{Nq}^{-1}\,.
\end{align}
The index $q$ enumerates different polynomials of the same degree.
The polynomials $P_{Nq}$ can be obtained as solutions of one-loop RG equation for twist$-3$
three quark operators. The expansion coefficients $\phi_{Nq}(\mu_R)$ 
are related to the nucleon matrix element of local three quark operators
\begin{align}\label{LocO}
\mathbb{O}^{t=3}_{Nq}(\mu_R)= P_{Nq}(\partial_z)\,\mathbb{O}_{3}({z},\mu_R)|_{z=0}.
\end{align}
Here, $\mathbb{O}_3(z,\mu_R)$ is the (renormalized) light-ray operator
\begin{align}\label{def:t3oper}
\mathbb{O}_{3}(z) = \epsilon^{ijk}u^{\downarrow i}_+(z_1)u^{\uparrow j}_+(z_2) d^{\downarrow
k}_+(z_3)\,.
\end{align}
Throughout this paper $z=\{z_1,z_2,z_3\}$ and from now on we do not show the scale
dependence.

The matrix element of the operator~(\ref{LocO}) can be parameterized as follows
\begin{align}\label{def:phi_Nq}
\vev{0|\mathbb{O}^{t=3}_{Nq}|N}=-\frac{i}2  N^{\downarrow}_+\,(-ipn)^{N+1}\,\phi_{Nq}\,.
\end{align}
The reduced matrix element $\phi_{Nq}$ can be expressed as a convolution integral
\begin{align}\label{}
\phi_{Nq}=\int \mathcal{D}x\,P^\dagger_{Nq}(x)\Phi_3(x)\,.
\end{align}
To the one-loop accuracy the reduced matrix elements $\phi_{Nq}(\mu)$ have  an autonomous scale
dependence. More details can be found in  Refs.~\cite{BDKM,Braun:2008ia}.

The expansion of the higher twist DAs has a similar form
\begin{align}\label{Fe}
\mathcal{F}(x)=\omega_{\mathcal{F}}(x)\sum_{N,q} C^{\mathcal{F}}_{Nq}\eta^{\mathcal{F}}_{Nq}(\mu_R)
\mathcal{P}^{\mathcal{F}}_{Nq}(x_1,x_2,x_3)+\ldots\,,
\end{align}
where dots stand for the contribution of quark-gluon operators which are irrelevant for
our purposes.
For the twist$-4$ DAs, $\mathcal{F}=\{\Phi_4,\Psi_4,\Xi_4\}$, the weight function takes the form
$\omega_{\mathcal{F}}(x)=\{x_1x_2,\ x_1x_3, \ x_2x_3\}$, respectively, and for the twist$-5$ DAs,
$\mathcal{F}=\{\Phi_5,\Psi_5,\Xi_5\}$ one finds
$\omega_{\mathcal{F}}(x)=\{x_3,\ x_2,\ x_1\}$. The expansion coefficients
$\eta^{\mathcal{F}}_{Nq}(\mu)$ are related to the nucleon matrix elements of the (multiplicatively renormalized)
local operators of  a {\it collinear} twist four and five, respectively. The corresponding polynomials
$\mathcal{P}^{\mathcal{F}}_{Nq}$ can be obtained by solving RG equations. (For the twist four
functions, this expansion was worked out in detail in Ref.~\cite{Braun:2008ia}).

Since DAs $\Phi_4$ and $\Psi_4$ have the \textit{collinear} twist four
they  receive contributions from the operators of both
the \textit{geometrical} twist four and three.
The part due to twist$-3$ operators is nothing else than the Wandzura-Wilczek contribution
\begin{align}\label{}
 \Phi_4=\Phi_4^{t=4}+\Phi_4^{WW},&&\Psi_4=\Psi_4^{t=4}+\Psi_4^{WW}.
 \end{align}
The chiral DA $\Xi_4$ does not receive contributions from twist$-3$ operators, i.e.
$\Xi_4=\Xi_4^{t=4}$. Conformal expansion for
the 
functions $\Phi_4^{WW},\,\,\Psi_4^{WW}$ was obtained in Ref.~\cite{Braun:2008ia}.

Quite similarly, the  twist$-5$ DAs can be represented as
\begin{align}\label{}
\Phi_5&=\Phi_5^{t=5}+\Phi_5^{WW},&&\Psi_5=\Psi_5^{t=5}+\Psi_5^{WW}\,,
\notag\\[2mm]
\Xi_5&=\Xi_5^{t=5}+\Xi_5^{WW}\,. &&
\end{align}
The functions $\Phi_5^{t=5}, \Psi_5^{t=5}$ and $\Xi_5^{t=5}$ contain the
contributions from the local operators of geometrical twist$-5$, while the WW parts
entail the contributions from the operators of geometrical twist three and four.

\section{Twist$-3$ contribution to $\Psi_4$.}\label{PF4}

The WW contribution to the DAs  $\Phi_4,\ \Psi_4$ was originally derived
in~\cite{Braun:2008ia}.
Here we  streamline the derivation and obtain some results which will be used
in the next section.

The nonlocal three-quark operator of twist$-3$ was defined in
Eq.~(\ref{def:t3oper}). Here we introduce two operators of collinear twist$-4$
\begin{subequations}
\begin{align}\label{E1}
\mathbb{O}_4({z},\mu)&=\epsilon^{ijk}
u_{-}^{\downarrow i }(z_1)  u_{+}^{\uparrow j}(z_2) d_{+}^{\downarrow k}(z_3)\,,
\\[2mm]
\label{E2}
\widetilde{\mathbb{O}}_4({z},\bar\mu)&=\epsilon^{ijk}
u_+^{\downarrow i }(z_1) u_{-}^{\uparrow j}(z_2) d_{+}^{\downarrow k}(z_3)\,.
\end{align}
\end{subequations}
The dependence of light-ray operators on the auxiliary spinors $\lambda,\bar\lambda$
is always implied and we will not display it explicitly.

The matrix element of the operator $\mathbb{O}_4({z},\mu)$ defines the DA $\Psi_4$, while
the nucleon matrix element of the operator $\widetilde{\mathbb{O}}_4({z},\bar\mu)$
vanishes because its helicity is equal to $3/2$. Nevertheless,
we consider this operator since it will be necessary for our analysis of twist$-5$ functions.

The nonlocal operators $\mathbb{O}_3({z}),\ \mathbb{O}_4({z},\mu),\
\widetilde{\mathbb{O}}_4({z},\bar\mu)$ transform in a proper way under the transformations of
the collinear $SL(2,R)$ subgroup of the conformal group.
The representation $T^{j}$ of the $SL(2,R)$ group ($j$ is called a conformal spin) is
defined by a transformation law
\begin{align}\label{}
[T^{j}(g^{-1})f](z)=\frac1{(cz+d)^{2j}} f\left(\frac{az+b}{cz+d}\right)\,,
\end{align}
where $g=\begin{pmatrix}a&b\\c&d\end{pmatrix}$ is a unimodular real matrix.
The generators of infinitesimal transformations $S_\pm, S_0$ have the form
\begin{align}\label{}
S_+=z^2\partial_z+2jz\,, && S_0=z\partial_z+j\,, && S_-=-\partial_z
\end{align}
and obey the standard $sl(2)$ commutation relations
\begin{align}\label{comrel}
[S_+,S_-]=2S_0, && [S_0,S_\pm]=\pm S_\pm.
\end{align}
The $'\pm'$ projections of  the quark field, $q_+(z)$ and $q_-(z)$, transform according to the
representations $T^{j=1}$ and $T^{j=1/2}$, respectively. Thus, the operator
$\mathbb{O}_3({z})$ transforms according to the tensor product of representations, $T^{1}\otimes T^{1}\otimes
T^{1}$, while the operators $\mathbb{O}_4({z},\mu)$ and
$\widetilde{\mathbb{O}}_4({z},\bar\mu)$ transform according to the tensor products
$T^{1/2}\otimes T^{1}\otimes
T^{1}$ and $T^{1}\otimes T^{1/2}\otimes
T^{1}$, respectively.

To the one loop accuracy (which we restrict  ourselves to),
the expansion of nonlocal operators in terms of local multiplicatively renormalized
operators reads as follows
\begin{align}\label{O3exp}
\mathbb{O}_3({z})=
\sum_{N,k,q} a_{Nk}\, S_+^k\Phi_{Nq}(\vec{z})\,
\,\partial_+^k\,\mathbb{O}_{Nq}^{t=3}.
\end{align}
Here $\partial_+=(n\partial)$ is the derivative along the $n$ direction.
The operator $S_+=S_{1,+}+S_{2,+}+S_{3,+}$ is the sum of one-particle generators.
(In what follows we will  assume that the conformal spins of the generators are
always determined by the transformation properties of the objects they act on.) The
coefficient $a_{Nk}$ in the expansion~(\ref{O3exp}) has the form
\begin{align}\label{}
a_{Nk}=\frac{\Gamma(2N+6)}{k!\Gamma(2N+6+k)}\,,
\end{align}
which follows immediately from the consistency equation
$\Big(\partial_++S_-\Big)\mathbb{O}_3({z})=0$.
The functions $\Phi_{Nq}(z)$ and $P_{Nq}(x)$ form a biorthogonal basis~\cite{Derkachov:1997qv,BDKM}
\begin{align}\label{}
P_{Nq}(\partial_z)\Phi_{N'q'}(z)\Big|_{z=0}=\delta_{NN'}\delta_{qq'}\,.
\end{align}
To the one loop accuracy  multiplicatively renormalized operators $\mathbb{O}_{Nq}^{t=3}$
are known to be the conformal operators that is they obey the following equation
\begin{align}\label{LW}
[\mathbf{K}_-,\mathbb{O}^{t=3}_{Nq}]=0,
\end{align}
where $\mathbf{K}_{-}=\bar n^\rho \mathbf{K}_\rho$ and $\mathbf{K}_\rho$ is the
generator of special conformal transformations. We note also that the functions
$\Phi_{Nq}(z)$ are shift invariant polynomials while the polynomials $P_{Nq}(x)$ obey
the equation
\begin{align}\label{}
\left(\sum_{k} x_k\partial_{x_i}^2+2j_k\partial_{x_k}\right)P_{Nq}(x)=0\,.
\end{align}
%
The conformal expansion for the twist$-4$ operators $\mathbb{O}_4(z,\mu)$ and
$\widetilde{\mathbb{O}}_4(z,\bar \mu)$ reads
\begin{subequations}\label{T4E}
\begin{align}
\mathbb{O}_4({z},\mu)&=
\sum_{N,k,q} b_{Nk}\, S_+^k\Psi_{Nq}({z})\,
\,\partial_+^k\,\mathbb{O}_{Nq}^{t=4}(\mu)\,,
\\
\widetilde{\mathbb{O}}_4({z},\bar\mu)&=
\sum_{N,k,q} b_{Nk}\, S_+^k\widetilde\Psi_{Nq}({z})\,
\,\partial_+^k\,\widetilde{\mathbb{O}}_{Nq}^{t=4}(\bar\mu)\,,
\end{align}
\end{subequations}
where
\begin{align}\label{}
b_{Nk}=\frac{\Gamma(2N+5)}{k!\Gamma(2N+5+k)}
\end{align}
and we tacitly assumed that the operator $S_+$
involves proper conformal spins that correspond to the  transformation properties of these
operators.
The operators $\mathbb{O}_{Nq}^{t=4}(\mu)$
($\widetilde{\mathbb{O}}_{Nq}^{t=4}(\bar\mu)$) are  the conformal (lowest weight) operators, {\it i.e.}
\begin{align}\label{LWC}
[ \mathbf{K}_-,\mathbb{O}_{Nq}^{t=4}(\mu)]~=~[\mathbf{K}_-,\widetilde{\mathbb{O}}_{Nq}^{t=4}(\bar\mu)]=0\,.
\end{align}
Among such lowest~-~weight operators there are  descendants of
the twist$-3$ operators, i.e. the operators of the collinear twist
four which can be expressed of terms of the 
operators~$\mathbb{O}^{t=3}_{Nq}$.

To construct them, we recall that operators $\mathbb{O}_{Nq}^{t=3}$ are homogeneous
functions of auxiliary spinors $\lambda$ and
$\bar\lambda$,
$\mathbb{O}_{Nq}^{t=3}\mapsto \mathbb{O}_{Nq}^{t=3}(\lambda,\bar\lambda)$.
It follows directly from the definition~(\ref{LocO}) that
$$
\mathbb{O}_{Nq}^{t=3}(a\lambda,\bar a\bar\lambda)=a^{N+2} \bar
a^{N+1}\mathbb{O}_{Nq}^{t=3}(\lambda,\bar\lambda).
$$
The operators $\mathbb{O}^{t=4}_{Nq}~ (\widetilde{\mathbb{O}}_{Nq}^{t=4}(\bar\mu))$ involve
one power of  the auxiliary spinor $\mu ~(\bar \mu)$ and
are  homogeneous polynomials in $\lambda,\bar\lambda$ of the degree $(N+1,N+1)$
($N+2, N$), respectively. For  brevity, we will not display $\lambda,\bar\lambda$ as the
arguments of the operators.

It is easy to check that the operators
\begin{subequations}\label{O41}
\begin{align}
\mathbb{O}^{t=4,(1)}_{Nq}(\mu)&=\frac1{N+2}(\mu\partial_\lambda)\,\mathbb{O}^{t=3}_{Nq}
\,,
\\
\widetilde{\mathbb{O}}^{t=4,(1)}_{Nq}(\bar\mu)&=\frac1{N+1}(\bar\mu\partial_{\bar\lambda})\,
\mathbb{O}^{t=3}_{Nq}
\end{align}
\end{subequations}
have the necessary degree of homogeneity. Taking into account Eq.~(\ref{LW}) and that
$\mathbf{K}_-=\frac12\mu^\alpha \mathbf{K}_{\alpha\dot\alpha}\bar\mu^{\dot\alpha}$ does
not depend on $\lambda,\ \bar\lambda$ one immediately  verifies that these operators
satisfy the lowest weight condition~(\ref{LWC}).
Hence they can enter the expansion~(\ref{T4E}).

The other set of the operators satisfying~(\ref{LWC}) involves commutators with the
momentum operator $\mathbf{P}$
\begin{subequations}\label{O42}
\begin{align}
\mathbb{O}^{t=4,(2)}_{N+1,q}(\mu
) &= \frac{1}{4(N+3)^2}\Biggl(i\Big[\mathbf{P}_{\mu\bar\lambda}\,,
\mathbb{O}^{t=3}_{Nq}
\Big]
\notag\\
&\quad-\frac{N+2}{2N+5}i\Big[\mathbf{P}_{\lambda\bar\lambda}\,,
\mathbb{O}^{t=4,(1)}_{Nq}(\mu)
\Big]\Biggr),
\\
\widetilde{\mathbb{O}}^{t=4,(2)}_{N+1,q}(\bar\mu
)&=\frac{1}{4(N+3)(N+4)}\Biggl(i\Big[\mathbf{P}_{\lambda\bar\mu}\,,
\mathbb{O}^{t=3}_{Nq}
\Big]
\notag\\
&\quad -\frac{N+1}{2N+5}i\Big[\mathbf{P}_{\lambda\bar\lambda}\,,
\widetilde{\mathbb{O}}^{t=4,(1)}_{Nq}(\bar\mu)
\Big]\Biggr).
\end{align}
\end{subequations}
Here, $\mathbf{P}_{\mu\bar\lambda}\equiv \mu^\alpha
\mathbf{P}_{\alpha\dot\alpha}\bar\lambda^{\dot\alpha}$ and similar for others. Obviously,
these operators are homogeneous polynomials in $\lambda,\bar\lambda$ of the required
degree. Making use of the commutation relations for the generators of conformal algebra
(see Appendix~\ref{app:conf}) and taking into account Eq.~(\ref{LW}), one can easily check
that the above operators  satisfy the lowest weight condition~(\ref{LWC}).

We would like to stress that the twist$-3$ operator with $N$ derivatives,
$\mathbb{O}^{t=3}_{Nq}$, gives rise to two twist$-4$ operators with $N$ and $N+1$
derivatives, respectively. Such a phenomenon was observed in Ref.~\cite{Ball:1998sk}
where it was
shown that this effect is related to the spin rotation of hadron in the rest frame.

The operators of geometric twist$-4$ contain the factor
$(\mu\lambda)$ or $(\bar\lambda\bar\mu)$. Indeed, to construct the operator of geometric
twist$-4$ one has to antisymmetrize a pair of indices and symmetrize all others.
After contraction with auxiliary spinors $\mu(\bar\mu)$ and $\lambda,\bar\lambda$ the antisymmetrized pair
produces the factor $(\mu\lambda)$ ($(\bar\lambda\bar\mu)$).

Our next task is to determine the functions, $\Psi_{Nq}^{(1)},\ \widetilde{\Psi}_{Nq}^{(1)}$ and
$\Psi_{Nq}^{(2)},\ \widetilde{\Psi}_{Nq}^{(2)}$ which
accompany the operators~(\ref{O41}),~(\ref{O42}) in the expansion~(\ref{T4E}). To do this
we note that replacing  $"-"\to "+"$ ($\mu\to\lambda$) in the definition~(\ref{E1}) for the nonlocal operator
$\mathbb{O}_4(z,\mu)$ one gets the twist$-3$ operator~$\mathbb{O}_3(z)$.
%
%
Formally, it can be written as
\begin{align}\label{EQ4}
\lambda^\alpha\dfrac{\partial}{\partial\mu^\alpha}\mathbb{O}_4(z,\mu)
&\equiv(\lambda\partial_\mu)\mathbb{O}_4(z,\mu)=
\mathbb{O}_3(z)\,,
\notag\\
\bar\lambda^{\dot\alpha}\dfrac{\partial}{\partial\bar\mu^{\dot
\alpha}}\widetilde{\mathbb{O}}_4(z,\bar\mu)
&\equiv(\bar\lambda\partial_{\bar\mu})\widetilde{\mathbb{O}}_4(z,\bar\mu)
=
\mathbb{O}_3(z)\,.
\end{align}
First of all, we stress that the operation $(\lambda\partial_\mu)$,
($(\bar\lambda\partial_{\bar\mu})$) kills all operators of the geometric twist$-4$ in the
expansion~(\ref{T4E}). Indeed, as was discussed above, they contain the factor
$(\mu\lambda)$ and $(\lambda\partial_\mu)(\mu\lambda)=(\lambda,\lambda)=0$.
Then, taking into account
\begin{align}\label{}
(\lambda\partial_\mu)\mathbb{O}^{t=4,(1)}_{Nq}(\mu)=
(\bar\lambda\partial_{\bar\mu})\widetilde{\mathbb{O}}^{t=4,(1)}_{Nq}(\bar\mu)
=\mathbb{O}^{t=3}_{Nq}
\end{align}
and
\begin{align}\label{}
(\lambda\partial_\mu)\mathbb{O}^{t=4,(2)}_{N+1,q}(\mu)&=
(\bar\lambda\partial_{\bar\mu})\widetilde{\mathbb{O}}^{t=4,(2)}_{N+1,q}(\bar\mu)=
\notag\\
&=
\frac1{2(N+3)(2N+5)}\partial_+\mathbb{O}^{t=3}_{N+1,q} \,,
\end{align}
we can bring Eqs.~(\ref{EQ4}) into the following form
\begin{align}\label{}
\sum_{N,k,q}\Biggl\{ \Big(a_{Nk}S_+^k\Phi_{Nq}({z})-b_{Nk}\,\widetilde S_+^k\Psi^{(1)}_{Nq}({z})\Big)
\,\partial_+^k\,\mathbb{O}_{Nq}^{t=3}
\notag\\
-\frac{b_{N+1k}}{2(N+3)(2N+5)} \widetilde S_+^k\Psi^{(2)}_{N+1q}({z})\,
\,\partial_+^{k+1}\,\mathbb{O}_{Nq}^{t=3}\Biggl\}=0
\end{align}
and similar for the second equation. The generators acting on the polynomials $\Phi_{Nq},
\,\Psi_{Nq}^{(a)}$ correspond to different conformal spins, $S_+=S_+^{(111)}$ and
$\widetilde S_+=S_+^{(\frac1211)}$ since the
operators $\mathbb{O}_3$ and $\mathbb{O}_4$ transform according to
different representations of the  $SL(2,R)$ group.
The coefficients at $\partial_+^k\mathbb{O}_{Nq}^{t=3}$ have to vanish identically for
arbitrary $k$.
To fix the functions $\Psi^{(1)}_{Nq}$ and $\Psi^{(2)}_{Nq}$  it is sufficient to
consider the equations for $k=0,1$. Indeed,  the equation for $k=0$  results in
\begin{align}\label{PW41}
\Psi^{(1)}_{Nq}({z})=\Phi_{Nq}({z})\,,
\end{align}
while the equation for  $k=1$ gives
\begin{align}\label{PW42}
\Psi^{(2)}_{N+1q}({z})=\Big[(2N+5) S_+
-(2N+6)\widetilde S_+
\Big]\Phi_{Nq}({z}).
\end{align}
Using the representation~(\ref{A1}) for the function $S_+^k\Psi_{Nq}^{(i)}$,
one can check that all equations for $k>1$ are satisfied provided $\Psi_{Nq}^{(i)}$ are
given by Eqs.~(\ref{PW41}),~(\ref{PW42}).

For the  functions $\widetilde\Psi_{Nq}^{(a)}$, the equations take the form
\begin{align}\label{PWT4}
\widetilde{\Psi}^{(1)}_{Nq}({z})&=\Phi_{Nq}({z})\,,
\\
\widetilde{\Psi}^{(2)}_{N+1q}({z})&=\Big[(2N+5)S_+^{(111)}-(2N+6)S_+^{(1\frac121)}\Big]\Phi_{Nq}({z}).
\notag
\end{align}
Note that the coefficient functions $\Phi_{Nq} (\Psi_{Nq})$ which accompany
the conformal (lowest weight) operators in the expansion of nonlocal operators are the shift
invariant polynomials, $S_-\Phi_{Nq}(z)=0$ ($S_-\Psi_{Nq}(z)=0$) (see Refs.~\cite{BDKM,Derkachov:1997qv} for details).
One can easily check that the polynomials~(\ref{PW42}), (\ref{PWT4}) indeed satisfy this condition.

Thus one gets the following expression
for the contributions of the descendants of the twist$-3$ operators to the
light-ray operators $\mathbb{O}_4(z)$ (and similarly for $\widetilde{\mathbb{O}}_4(z)$)
\begin{align}\label{PsiO4}
\mathbb{O}^{WW}_4(z)&=
\sum_{N,k,q} b_{Nk}\, S_+^k\Biggl\{\Psi^{(1)}_{Nq}({z})\,
\,\partial_+^k\,\mathbb{O}^{t=4,(1)}_{Nq}(\mu)
\notag\\
&\quad+
\Psi^{(2)}_{Nq}({z})\,
\,\partial_+^k\,\mathbb{O}^{t=4,(2)}_{Nq}(\mu)\Biggr\}.
\end{align}
In order to find $\Psi_4^{WW}(x)$ we take the nucleon matrix elements of both sides of
Eq.~(\ref{PsiO4}). By definition
\begin{align}\label{WWlhs}
\vev{0|\mathbb{O}^{WW}_4(z)|P}&=
\frac{1}4(\mu\lambda)  m_NN^{\uparrow}_+\int \mathcal{D}x\,e^{-i(pn)\sum z_k x_k}\notag\\
&\quad\times
\Psi^{WW}_4(x_2,x_1,x_3).
\end{align}
In its turn, for the  matrix elements of the operators $\mathbb{O}^{t=4,(1)}_{Nq}(\mu)$,
$\mathbb{O}^{t=4,(2)}_{Nq}(\mu)$ one derives
\begin{align}\label{Mel4}
\vev{0|\mathbb{O}^{t=4,(1)}_{Nq}|P}&=\frac1{4}(\mu\lambda) m_N N^{\uparrow}_+
\frac{(-ipn)^N\phi_{Nq}}{N+2}\,,
\\
\vev{0|\mathbb{O}^{t=4,(2)}_{N+1q}|P}&=-\frac1{8}(\mu\lambda) m_N N^{\uparrow}_+
\frac{(-ipn)^{N+1}\phi_{Nq}}{(N+3)^2(2N+5)}.\notag
\end{align}
Here we take into account that nucleon has zero transverse momentum,   $p_{\mu\bar\lambda}=p_{\lambda\bar\mu}=0$,
and use the equation of motion relation $2(pn)N^{\downarrow}_-=-(\mu\lambda) m_N N^\uparrow_+$.

Using Eqs.~(\ref{Mel4}) and  the summation formula Eq.~(\ref{A4}), one can bring the matrix element of the
rhs of Eq.~(\ref{PsiO4}) to the form
\begin{multline}\label{WWrhs}
\frac1{4}(\mu\lambda) m_N N^{\uparrow}_+ \sum_{Nq}\Gamma(2N+5)\,\phi_{Nq}\int \mathcal{D} x \, x_2 x_3
\,e^{-i(pn)\sum z_k x_k}
\\
\times\Biggl\{\frac1{N+2}{P}_{N,q}^{(1)}(x)-\frac{1}{N+3}P^{(2)}_{N+1,q}(x)\Biggr\}\,,
\end{multline}
where the polynomials ${P}_{N,q}^{(1)}(x)$, ${P}_{N+1,q}^{(2)}(x)$ are given by the $sl(2)$
Fourier transform
\begin{align}\label{}
{P}_N^{(k)}(x)=\VEV{\text{e}^{\sum_{i=1}^3 x_i z_i}|\Psi^{(k)}_{Nq}}_{\frac12 1 1}\,.
\end{align}
Here $\vev{*,*}$ stands for the $sl(2)$ invariant scalar product that is defined in
Appendix~\ref{App:UR}.
We have to express these polynomials in terms of $P_{Nq}$ which enter the expansion for
twist$-3$ nucleon DA, $\Phi_3(x)$. It follows from Eqs.~(\ref{DA3}),~(\ref{Phi3}),~(\ref{O3exp})
that
\begin{align}\label{}
c_{Nq}P_{Nq}(x)={\Gamma(2N+6)}\VEV{{e}^{\sum_{i=1}^3 x_i z_i}|\Phi_{Nq}}_{1 1 1}\,.
\end{align}
Taking into account Eq.~(\ref{SSrel}), one gets for $P_{Nq}^{(1)}$,
\begin{align}\label{PNq1}
P_{Nq}^{(1)}(x)=r_{Nq}\,\partial_{x_1} x_1P_{Nq}(x)\,,
\end{align}
where $r_{Nq}={c_{Nq}}/{\Gamma(2N+6)}$.
Since the generators $S_+^{j}$ and $S_-^{j}$ (here $j$ is
multiindex $j=(j_1,j_2,j_3)$) are conjugated with respect to the corresponding scalar product,
$\vev{\Psi|S_+^{j}\Phi}_j=-\vev{S_-\Phi|\Psi}_j$, we obtain
\begin{multline}\label{PNq2}
P_{N+1,q}^{(2)}(x)=
r_{Nq}\,\Big[(2N+5)-x_{123}\partial_{x_1}\Big]x_1P_{Nq}(x)\,,
\end{multline}
where $x_{123}=x_1+x_2+x_3$.
Inserting~(\ref{PNq1}),~(\ref{PNq2}) into~(\ref{WWrhs}) and comparing the result with~(\ref{WWlhs}),
we derive the following expression
\begin{multline}\label{PsiWW4}
\Psi_4^{WW}(x)=-\sum_{Nq}\frac{c_{Nq}\phi_{Nq}}{(N+2)(N+3)}
\\
\times\left[N+2-\partial_{x_2}\right]x_1x_2x_3\,P_{Nq}(x_2,x_1,x_3)\,
\end{multline}
that agrees with~\cite{Braun:2008ia}. The expression for $\Phi^{WW}_4$ does not require a
new calculation. It can be obtained from~(\ref{PsiWW4}) by a permutation of arguments, see
 sect.~\ref{Results}.
\vskip 5mm

\section{Twist$-3$ contribution to $\Phi_5^{WW}$.}\label{PF5}

The function  $\Phi_5$ receives the WW contributions from the operators of the geometric
twist three and four. We discuss only the twist$-3$ contribution because it is a bit involved.
The calculation of twist$-4$ part is straightforward and follows the lines of the previous section.
The  function $\Psi^{WW}_5$  can be obtained from  $\Phi_5^{WW}$ by a simple
permutation of arguments.

The distribution amplitude $\Phi_5$ is determined by the nucleon matrix element of the light-ray operator
\begin{align}\label{O5a}
\mathbb{O}_5(z)=\epsilon^{ijk} u_-^{\downarrow i}(z_1) u_-^{\uparrow j}(z_2) d_{+}^{\downarrow k}(z_3)\,.
\end{align}
This operator transforms according to the tensor product of $SL(2,R)$ representations,
$T^{1/2}\otimes T^{1/2}\otimes T^{1}$.
The expansion of the light-ray operator~(\ref{O5a}) over local  operators reads
\begin{align}\label{O5}
\mathbb{O}_5(z)=\sum_{N,q,k} d_{Nk} \,S_+^k\Upsilon_{Nq}(z)\,\partial_+^k
\mathbb{O}^{t=5}_{Nq}\,.
\end{align}
We emphasize that
the three particle generator $S_+$ in this expression carries the conformal spins
corresponding to the transformation properties of the operator~$\mathbb{O}_5(z)$,  $j_1=j_2=1/2$, $j_3=1$.
The coefficient $d_{Nk}$ have the following form
\begin{align}\label{dnk}
d_{Nk}=\frac{\Gamma(2N+4)}{k!\Gamma(2N+4+k)}\,.
\end{align}
Again, among the lowest weight operators $\mathbb{O}^{t=5}_{Nq}$ contributing to~(\ref{O5}) there are
descendants of the  twist$-3$ and twist$-4$ operators $\mathbb{O}^{t=3}_{Nq}$,\
$\mathbb{O}^{t=4}_{Nq}$. The descendants of twist$-3$ operators can be chosen as follows:

(i) the first descendant of $\mathbb{O}_{Nq}^{t=3}$ takes the form
\begin{align}\label{O51}
\mathbb{O}_{Nq}^{t=5,(1)}(\mu,\bar\mu)=\frac1{(N+1)(N+2)} 
(\mu\partial_\lambda)(\bar\mu\partial_{\bar\lambda})
\mathbb{O}^{t=3}_{Nq}\,,
\end{align}
(ii) the two more are related to the operators defined in~(\ref{O42})
\begin{align}\label{O523}
{\mathbb{O}}^{t=5,(2)}_{N+1,q}(\mu,\bar\mu)&=  \frac1{N+2}
(\bar\mu\partial_{\bar\lambda})\,{\mathbb{O}}^{t=4,(2)}_{N+1,q}(\mu)\,,
\notag\\[2mm]
{\mathbb{O}}^{t=5,(3)}_{N+1,q}(\mu,\bar\mu) &=  \frac1{N+3}
(\mu\partial_{\lambda})\,
\widetilde{\mathbb{O}}_{N+1,q}^{t=4,(2)}(\bar\mu)\,,
\end{align}
(iii) the last two operators are
\begin{align}\label{O545}
{\mathbb{O}}^{t=5,(4)}_{N+1,q}(\mu,\bar\mu) &= i[\mathbf{P}_{\mu\bar\mu},\mathbb{O}^{t=3}_{Nq}]\,,
\notag\\[2mm]
{\mathbb{O}}^{t=5,(5)}_{N+2,q}(\mu,\bar\mu)&=
          \frac1{2(N+3)}\Biggl\{[i\mathbf{P}_{\lambda\bar\lambda}[i\mathbf{P}_{\mu\bar\mu},\mathbb{O}^{t=3}_{Nq}]]
\notag\\
&\quad-
(\mu\partial_\lambda)[i\mathbf{P}_{\lambda\bar\lambda},
[i\mathbf{P}_{\lambda\bar\mu},\mathbb{O}^{t=3}_{Nq}]]
\notag\\
&\quad-
(\bar\mu\partial_{\bar\lambda})[i\mathbf{P}_{\lambda\bar\lambda},
[i\mathbf{P}_{\mu\bar\lambda},\mathbb{O}^{t=3}_{Nq}]]
\notag\\
&\quad+\frac{(\mu\partial_\lambda)(\bar\mu\partial_{\bar\lambda})}{
(2N+7)}[i\mathbf{P}_{\lambda\bar\lambda}[i\mathbf{P}_{\lambda\bar\lambda}
\mathbb{O}^{t=3}_{Nq}]]\Biggr\}
\notag\\
&\quad+[i\mathbf{P}_{\mu\bar\lambda}[i\mathbf{P}_{\lambda\bar\mu},\mathbb{O}^{t=3}_{Nq}]]\,.
\end{align}
All these operators are the lowest weight (conformal) operators,
$[\mathbf{K}_-,\mathbb{O}^{t=5,(a)}_{Nq}]=0$, $a=1,\ldots,5$.
This fact is obvious for the first four operators while for the last one
it can be checked with a help of the commutation relations given in Appendix~\ref{app:conf}.

The next step is to determine the coefficient functions $\Upsilon_{Nq}^{(a)}$ in
the expansion~(\ref{O5}). To this end we consider the following two equations
\begin{subequations}\label{T5E}
\begin{align}\label{T5Ea}
(\bar\lambda\partial_{\bar\mu})\mathbb{O}_{5}(z)&=\mathbb{O}_4(z)\,,
\\
\label{T5Eb}
(\lambda\partial_{\mu})\mathbb{O}_{5}(z)&=\widetilde{\mathbb{O}}_4(z)\,.
\end{align}
\end{subequations}
Let us substitute the expansions~(\ref{O5}) and (\ref{T4E}) into Eqs.~(\ref{T5E}).
The derivatives $(\lambda\partial_\mu)\mathbb{O}^{t=5,(a)}_{Nq}$,
$(\bar\lambda\partial_{\bar\mu})\mathbb{O}^{t=5,(a)}_{Nq}$ of the local operators
can be expressed in terms of the operators~(\ref{O41}),~(\ref{O42}) and their $\partial_+$
derivatives. For the sake of completeness, we collect the corresponding expressions in
Appendix~\ref{App:list}. Note that the derivatives
$(\bar\lambda\partial_{\bar\mu}),(\lambda\partial_{\mu})$
annihilate all operators of geometric twist$-5$ by the same reason as discussed in
sect.~\ref{PF4}.

As a result Eqs.~(\ref{T5E}) take the form:
\begin{align}\label{ex1}
\sum_{Nq k}\left(C_{Nqk}^{(1)} \partial_+^k \mathbb{O}_{Nq}^{t=4,(1)}
+C_{Nqk}^{(2)} \partial_+^k \mathbb{O}_{N+1,q}^{t=4,(2)}
\right)=0\,,
\notag\\
\sum_{Nq k}\left(\widetilde C_{Nqk}^{(1)} \partial_+^k \widetilde{\mathbb{O}}_{Nq}^{t=4,(1)}
+\widetilde C_{Nqk}^{(2)} \partial_+^k \widetilde{\mathbb{O}}_{N+1,q}^{t=4,(2)}
\right)=0\,.
\end{align}
The coefficients $C_{Nqk}^{(i)}$,  $\widetilde C_{Nqk}^{(i)}$ are given by some linear combinations of
the polynomials $\Upsilon_{Nq}^{(a)}$ . Since the operators
$\partial_+^k \mathbb{O}_{Nq}^{t=4,(a)},\
\partial_+^k \widetilde{\mathbb{O}}_{N+1,q}^{t=4,(a)}
$ are independent all coefficients $C_{Nqk}^{(i)}$, $\widetilde C_{Nqk}^{(i)}$ have to vanish.
This requirement results in
an infinite number of equations on the $\Upsilon_{Nq}-$functions.
Only a few of them, however, are independent. To fix the functions
$\Upsilon_{Nq}^{(a)}$, it is sufficient to consider  the equations $C_{Nqk}^{(i)}=0$, $\widetilde C_{Nqk}^{(i)}=0 $
for $k=0,1$.  All other equations will be satisfied automatically that can be checked with the help of Eq.~(\ref{A1}).

The equations $C_{Nq,k=0}^{(1)}=0$ and $\widetilde C_{Nq,k=0}^{(1)}=0 $
are equivalent each other and result in
\begin{align}\label{E51}
\Upsilon_{Nq}^{(1)}(z)=\Phi_{Nq}(z)\,.
\end{align}
Next, the equations $C_{Nq,k=1}^{(1)}=0$, $\widetilde C_{Nq,k=1}^{(1)}=0 $
and $C_{Nq,k=0}^{(2)}=0$, $\widetilde C_{Nq,k=0}^{(2)}=0 $
take the form
\begin{align}\label{E52}
\sum_{a=2}^{4} B_N^{a}\,\Upsilon_{N+1,q}^{(a)}(z)&=\left(\frac{S_+^{(\frac12 1 1)}}{2N+5}-
\frac{S_+^{(\frac12\frac121)}}{2N+4}\right)
\Phi_{Nq}(z),
\notag\\
\sum_{a=2}^{4} \widetilde B_N^{a}\,\Upsilon_{N+1,q}^{(a)}(z)&=\left(\frac{S_+^{(1\frac12  1)}}{2N+5}-
\frac{S_+^{(\frac12\frac121)}}{2N+4}\right)
\Phi_{Nq}(z)\,,
\notag\\
\sum_{a=2}^{4} A_N^{a}\,\Upsilon_{N+1,q}^{(a)}(z)&=\Psi_{N+1,q}^{(2)}(z)\,,
\notag\\
\sum_{a=2}^{4} \widetilde A_N^{a}\,\Upsilon_{N+1,q}^{(a)}(z)&=\widetilde\Psi_{N+1,q}^{(2)}(z)\,.
\end{align}
The coefficients $A_N^{a},\ \widetilde A_N^{a},\ B_N^{a},\ \widetilde B_N^{a}$ can be found in
Appendix~\ref{App:list}, Eq.~(\ref{ABAB}). These equations are sufficient
to fix the functions $\Upsilon_{N+1,q}^{(a)}$, $a=2,3,4$. In fact, it is sufficient to
consider any three of them, the last equation provides a consistency check.

The remaining function $\Upsilon_{N+2,q}^{(5)}$ can be determined from the equation $C_{Nq,k=1}^{(2)}=0$
which implies
\begin{align}\label{}
C_N\Upsilon_{N+2,q}^{(5)}(z)=\left(\frac{S_+^{(\frac12 1 1)}}{2N+7}
-\frac{S_+^{(\frac12\frac121)}}{2N+6}\right)\Psi_{N+1,q}^{(2)}\,,
\end{align}
where the coefficient $C_N$ is given in Eq.~(\ref{C_N}).
The functions $\Upsilon_{Nq}^{(a)}$ are shift invariant, $S_-\Upsilon_{Nq}^{(a)}=0$, as they should be.

To proceed further, we need to calculate the $sl(2)$ Fourier transform of the functions
$\Upsilon_{Nq}^{(a)}$
\begin{align}\label{}
\mathcal{P}_{N,q}^{(a)}(x)=\VEV{e^{\sum_{k=1}^3 x_k z_k}|\Upsilon_{Nq}^{(a)}}_{\frac12\frac121}\,.
\end{align}
Using the method described in the previous section we obtain
\begin{eqnarray}
\label{}
\mathcal{P}_{N,q}^{(1)}(x)\hspace{-.3cm}&&= r_{Nq}\,\partial_{x_1}\partial_{x_2} x_1 x_2 P_{Nq}(x)\,,
\notag\\
\mathcal{P}_{N+2,q}^{(5)}(x)\hspace{-.3cm}&&= r_{Nq}\, \varkappa_N\,(2N+6-x_{123}\partial_{x_2})
\notag\\
&&\quad\times
(2N+5-x_{123}\partial_{x_1})\, x_1 x_2\, P_{Nq}(x)\,,
\end{eqnarray}
where, again, $x_{123}=x_1+x_2+x_3$ and
\begin{align}\label{}
\varkappa_N=(4(N+3)(N+4)(2N+5)(2N+6))^{-1}\,.
\end{align}
The expressions for the polynomials $\mathcal{P}_{N+1,q}^{(2,3,4)}$ are a bit more
complicated
\begin{align}\label{}
\mathcal{P}_{N+1,q}^{(2)}(x)&=r_{Nq}\frac{2N+5}{(N+1)}
\Biggl\{-\frac{3N+5}{2(N+2)}x_{123}\partial_{x_1}\partial_{x_2}
\notag\\
&\quad+(N+1)\partial_{x_1}+2(N+2)\partial_{x_2}\Biggr\}\,x_1x_2\,P_{Nq}(x)\,,
\notag\\
\mathcal{P}_{N+1,q}^{(3)}(x)&=r_{Nq}\frac{(2N+5)(N+3)}{(N+1)(N+2)}
\Biggl\{-\frac{3N+4}{2(N+2)}x_{123}\partial_{x_1}\partial_{x_2}
\notag\\
&\quad+2(N+1)\partial_{x_1}+(N+2)\partial_{x_2}\Biggr\}\,x_1x_2\,P_{Nq}(x)\,,
\notag\\
\mathcal{P}_{N+1,q}^{(4)}(x)&=\frac{r_{Nq}}{2(N+1)(N+2)}\Biggl\{\frac12\frac{2N+3}{N+2}x_{123}\partial_{x_1}\partial_{x_2}
\notag\\
&\quad
-(N+1)\partial_{x_1}-(N+2)\partial_{x_2}\Biggr\}\, x_1 x_2\,P_{Nq}(x)\,.
\end{align}
Next, we represent the nucleon matrix element of the operators $\mathbb{O}_{Nq}^{t=5,(a)}$
in the form
\begin{align}\label{}
\vev{0|\mathbb{O}_{Nq}^{t=5,(a)}|P}=-\frac1{4}m_N(\mu\lambda)N^{\uparrow}_-\,(-ipn)^{N} \, \Theta_{Nq}^{(a)}\,.
\end{align}
Taking into account Eqs.~(\ref{O51}),~(\ref{O523}),~(\ref{O545})
and Eq.~(\ref{def:phi_Nq}),
one obtains for the reduced matrix elements $\Theta_{Nq}^{(a)}$
\begin{align}\label{}
\Theta_{Nq}^{(1)}&=\frac{\phi_{Nq}}{(N+2)}\,,
\notag\\
\Theta_{N+1,q}^{(2)}&=\frac{(N+4)\phi_{Nq}}{2(N+2)(N+3)^2(2N+5)}\,,
\notag\\
\Theta_{N+1,q}^{(3)}&=\frac{\phi_{Nq}}{2(N+3)^2(2N+5)}\,,
\notag\\
\Theta_{N+1,q}^{(4)}&=2\,\phi_{Nq}\,,
\notag\\
\Theta_{N+2,q}^{(5)}&=-\frac{2(N+4)\phi_{Nq}}{(N+3)(2N+7)}\,.
\end{align}
Using the summation formula~(\ref{A4}), one obtains for the Wandzura-Wilczek
contribution to the matrix element of the operator~$\mathbb{O}_5(z)$
\begin{align}\label{}
\vev{0|\mathbb{O}_{5}(z)|P}&=-\frac1{4}m_N(\mu\lambda)N^{\uparrow}_-\sum_{Nq} \int\mathcal{D}x\, x_3\,
e^{-ip_+\sum z_i x_i}\times
\notag\\
&\quad
\Biggl\{
\zeta_{N}\Theta_{Nk}^{(1)}\mathcal{P}_{Nq}^{(1)}(x)
 +\zeta_{N+2}\Theta_{N+2k}^{(5)}\mathcal{P}_{N+2,q}^{(5)}(x)
 \notag\\
 &\quad+
\zeta_{N+1}\sum_{a=2}^4\Theta_{N+1,q}^{(a)}\mathcal{P}_{N+1,q}^{(a)}(x)
\Biggr\}\,,
\end{align}
where $\zeta_N=\Gamma(2N+4)$. Substituting the explicit expressions for the polynomials
$\mathcal{P}_{Nq}$ and the reduced matrix elements $\Theta_{Nq}$, one finds after some algebra
\begin{align}\label{O5exp1}
\vev{0|\mathbb{O}^{(5)}(z)|P}&=-\frac1{4}m_N(\mu\lambda)N^{\uparrow}_-\int\mathcal{D}x\,
e^{-i(pn)\sum x_kz_k}
\notag\\
&\quad\times
\sum_{Nq}\frac{c_{Nq}\phi_{Nq}(\mu)}{(N+2)(N+3)} \mathcal{B}_{Nq}(x)\,,
\end{align}
where
\begin{align}\label{}
\mathcal{B}_{Nq}(x)&=\biggl[(N+2-\partial_{x_1})(N+1-\partial_{x_2})
\notag\\
&\quad
-(N+2)^2\biggr]x_1x_2x_3 P_{Nq}(x)\,.
\end{align}
Comparing~(\ref{O5exp1}) with the definition~(\ref{DA5}), one gets
\begin{align}\label{Phi5WW}
\Phi_5^{WW}(x)=\sum_{Nq}\frac{c_{Nq}\phi_{Nq}(\mu)}{(N+2)(N+3)} \mathcal{B}_{Nq}(x)\,.
\end{align}
The expression for $\Psi_5^{WW}(x)$ is given in sect.~\ref{Results} and can be restored
from~(\ref{Phi5WW}) by a simple permutation of arguments.
The calculation of the WW contribution from twist$-4$ operators to the DAs $\Phi_5$,
$\Psi_5$ and $\Xi_5$ is straightforward and we only present the final result in sect.~\ref{Results}.

\section{Results}\label{Results}

In this section we collect all results for the nucleon DAs. The conformal expansion for twist$-3$ and
genuine (geometric) twist$-4$ DAs reads
\begin{align}\label{defDA}
\Phi_3(x)&=x_1x_2x_3\sum_{Nq} c_{Nq}\,\phi_{Nq}(\mu_R)\, P_{Nq}(x)\,,
\notag\\
\Phi^{t=4}_4(x)&=x_1 x_2\sum_{Nq} A_{Nq}\,\eta_{Nq}(\mu_R)\, R_{Nq}(x)\,,
\notag\\
\Psi^{t=4}_4(x)&=x_1 x_3\sum_{Nq} \widetilde A_{Nq}\,\eta_{Nq}(\mu_R)\, \widetilde R_{Nq}(x)\,,
\notag\\
\Xi_4(x)&=x_2x_3\sum_{Nq} B_{Nq}\,\xi_{Nq}(\mu_R)\, \Pi_{Nq}(x)\,.
\end{align}
Explicit expressions for the first few functions $P_{Nq},\, R_{Nq},\, \widetilde
R_{Nq},\, \Pi_{Nq}$ can be found in Ref.~\cite{Braun:2009vc}.
The Wandzura-Wilczek contributions to  the twist four DAs $\Phi_4$ and $\Psi_4$
take the form~\cite{Braun:2009vc}
\begin{align}
\Phi_4^{WW}(x)&=-\sum_{Nq}\frac{c_{Nq}\phi_{Nq}(\mu_R)}{(N+2)(N+3)}\left[N+2-\partial_{x_3}\right]
\notag\\
&\quad\times x_1x_2x_3\,P_{Nq}(x_1,x_2,x_3)\,,
\notag\\[2mm]
\Psi_4^{WW}(x)&=-\sum_{Nq}\frac{c_{Nq}\phi_{Nq}(\mu_R)}{(N+2)(N+3)}\left[N+2-\partial_{x_2}\right]
\notag\\
&\quad\times x_1x_2x_3\,P_{Nq}(x_2,x_1,x_3)\,.
\end{align}
The WW contributions to the twist five DAs $\Phi_5$ and $\Psi_5$ from the twist$-3$ operators
read
\begin{align}\label{Phi5WW_3}
\Phi_5^{WW_3}(x)&=\sum_{Nq}\frac{c_{Nq}\,\phi_{Nq}(\mu)}{(N+2)(N+3)} \biggl[
(N+2-\partial_{x_1})(N+1-\partial_{x_2})
\notag\\
&\quad
-(N+2)^2
\biggr]x_1x_2x_3 P_{Nq}(x_1,x_2,x_3)\,,
\notag\\
\Psi_5^{WW_3}(x)&=\sum_{Nq}\frac{c_{Nq}\,\phi_{Nq}(\mu)}{(N+2)(N+3)} \biggl[
(N+2-\partial_{x_3})(N+1-\partial_{x_1})
\notag\\
&\quad -(N+2)^2
\biggr]x_1x_2x_3 P_{Nq}(x_2,x_1,x_3)\,,
\end{align}
The WW contribution from the geometric twist$-4$ operators takes the form
\begin{align}\label{}
\Phi_5^{WW_4}(x)&=\sum_{Nq}\frac{\widetilde A_{Nq}\,\eta_{Nq}(\mu_R)}{(N+1)(N+3)}
\left[N+1-\partial_{x_2}\right]
\notag\\
&\quad\times
x_2 x_3\, \widetilde R_{Nq}(x_2,x_1,x_3)\,,
\notag\\[2mm]
\Psi_5^{WW_4}(x)&=\sum_{Nq}\frac{A_{Nq}\,\eta_{Nq}(\mu_R)}{(N+1)(N+3)}
\left[N+1-\partial_{x_1}\right]
\notag\\
&\quad\times
x_1 x_2 \, R_{Nq}(x_2,x_1,x_3)\,.
\end{align}
Finally, for the chiral DA $\Xi_5$, we obtain
\begin{align}\label{}
\Xi_5^{WW_4}(x)&=-\sum_{Nq}\frac{B_{Nq}\xi_{Nq}(\mu_R)}{(N+1)(N+3)}
\notag\\
&\quad\times
\Biggl\{\left[N+1-\partial_{x_3}\right]x_1 x_3 \Pi_{Nq}(x_2,x_1,x_3)
\notag\\
&\quad-\left[N+1-\partial_{x_2}\right] x_1 x_2
\Big(\Pi_{Nq}(x_3,x_2,x_1)
\notag\\
&\quad
+\Pi_{Nq}(x_3,x_1,x_2)\Big)\Biggr\}\,.
\end{align}
One can check that the  first moments of the WW contribution to the  twist$-5$ DAs
agree with the results of Refs.~\cite{Braun:2001tj,Braun:2006hz}.
\section*{Acknowledgements}
The authors are grateful to V.~M.~Braun for useful discussions.
This work was supported by the German Research
Foundation (DFG), grant BR2021/5-2, grant BR 2021/6-1 and
in part by the RFBR (grant 12-02-00613) and the Heisenberg-Landau Program.


\appendix
\renewcommand{\theequation}{\Alph{section}.\arabic{equation}}

\section*{Appendices}
\section{Conformal group}\label{app:conf}

In this Appendix we collect the commutators of the conformal group generators that are necessary
to construct the lowest weight (conformal) operators. The generators in the spinor
representations and their commutators can be found in Ref.~\cite{Braun:2009vc}.
We list below the relevant ones. Starting from
\begin{align*}\label{}
[i\mathbf{K}_{\alpha\dot\alpha},i\mathbf{P}^{\beta\dot\beta}
]=4
\left(\delta^{\beta}_{\alpha}\,\delta^{\dot\beta}_{\dot\alpha}\,i\mathbf{ D}+
\delta^{\dot\beta}_{\dot\alpha}\, {i\mathbf{M}_{\alpha}}^{\beta}
+\delta^{\beta}_{\alpha}\,i{\widebar{ \mathbf{M}}_{\dot\alpha}}^{\phantom{\dot\alpha}\dot\beta}
 \right)\,,
\end{align*}
where $\mathbf{D}$, $\mathbf{M}_{\alpha\beta}{\beta}$ and $ \widebar{
\mathbf{M}}_{\dot\alpha\dot\beta}$ are the generators of dilatations and Lorentz
rotations, respectively, one gets
\begin{align}\label{KPcomm}
[i\mathbf{K}_{\mu\bar\mu},i\mathbf{P}_{\lambda\bar\lambda}]&=
4i\Big((\mu\lambda)(\bar\lambda\bar\mu)\mathbf{D}-(\bar\lambda\bar\mu)\mathbf{M}_{\mu\lambda}-
(\mu\lambda)\bar{\mathbf{M}}_{\bar\lambda\bar\mu}\Big),
\notag\\
[i\mathbf{K}_{\mu\bar\mu},i\mathbf{P}_{\mu\bar\lambda}]&=-4(\bar\lambda\bar\mu)\,i\mathbf{M}_{\mu\mu}\,,
\notag\\
[i\mathbf{K}_{\mu\bar\mu},i\mathbf{P}_{\lambda\bar\mu}]&=-4(\mu\lambda)\,i\bar{\mathbf{M}}_{\bar\mu\bar\mu}\,,
\end{align}
where $\mathbf{P}_{\mu\bar\lambda}=\mu^\alpha \mathbf{P}_{\alpha\dot\alpha} \bar\lambda^{\dot\alpha}$, etc.
One also finds
\begin{align*}
[i\mathbf{M}_{\mu\lambda}, i\mathbf{P}_{\mu\bar\lambda}]=\frac{(\mu\lambda)}2 i\mathbf{P}_{\mu\bar\lambda},
&&
[i\bar{\mathbf{M}}_{\bar\mu\bar\lambda}, i\mathbf{P}_{\mu\bar\lambda}]=
\frac{(\bar\mu\bar\lambda)}2  i\mathbf{P}_{\mu\bar\lambda},
\notag\\
[i\mathbf{M}_{\mu\lambda}, i\mathbf{P}_{\lambda\bar\mu}]=\frac{(\mu\lambda)}2 i\mathbf{P}_{\lambda\bar\mu},
&&
[i\bar{\mathbf{M}}_{\bar\mu\bar\lambda}, i\mathbf{P}_{\lambda\bar\mu}]=
\frac{(\bar\mu\bar\lambda)}2  i\mathbf{P}_{\lambda\bar\mu},
\notag\\
[i\mathbf{M}_{\mu\lambda}, i\mathbf{P}_{\mu\bar\mu}]=\frac{(\mu\lambda)}2 i\mathbf{P}_{\mu\bar\mu},
&&
[i\bar{\mathbf{M}}_{\bar\mu\bar\lambda}, i\mathbf{P}_{\mu\bar\mu}]=
\frac{(\bar\mu\bar\lambda)}2 i\mathbf{P}_{\mu\bar\mu}.
\end{align*}
Let  $\mathbb{O}^{j\bar j}(\lambda,\bar\lambda)$ be an operator of Lorentz spin $j,\bar j$
at point $x=0$ with all indices contracted with spinors $\lambda,\bar \lambda$.
Such an operator transforms under Lorentz rotations as follows
\begin{align*}\label{}
i[\mathbf{M}_{\alpha\beta},\mathbb{O}^{j\bar j}(\lambda,\bar\lambda)]&=
-\frac12\left(\lambda_\alpha\frac{\partial}{\partial\lambda^\beta}
+\lambda_\beta\frac{\partial}{\partial\lambda^\alpha}\right)\mathbb{O}^{j\bar j}(\lambda,\bar\lambda),
\notag\\
i[\widebar{\mathbf{M}}_{\dot\alpha\dot\beta},\mathbb{O}^{j\bar j}(\lambda,\bar\lambda)]&=
-\frac12\left(\bar\lambda_{\dot\alpha}\frac{\partial}{\partial\bar\lambda^{\dot\beta}}
+\bar\lambda_{\dot\beta}\frac{\partial}{\partial\bar\lambda^{\dot\alpha}}\right)\mathbb{O}^{j\bar j}(\lambda,\bar\lambda).
\end{align*}
One easily finds
\begin{align}\label{}
i[\mathbf{M}_{\mu\lambda},\mathbb{O}^{j\bar j}(\lambda,\bar\lambda)]&=
-2j\,(\mu\lambda)\,\mathbb{O}^{j\bar j}(\lambda,\bar\lambda)\,,
\notag\\
i[\widebar{\mathbf{M}}_{\bar\mu\bar\lambda},\mathbb{O}^{j\bar j}(\lambda,\bar\lambda)]&=
-2\bar j\,(\bar\lambda\bar\mu)\,\mathbb{O}^{j\bar j}(\lambda,\bar\lambda)\,,
\notag\\
i[\mathbf{M}_{\mu\mu},\mathbb{O}^{j\bar j}(\lambda,\bar\lambda)]
&=
-(\mu\lambda)\,(\mu\partial_\lambda)\,\mathbb{O}^{j\bar j}(\lambda,\bar\lambda)\,,
\end{align}
etc. The above equations are useful to check the lowest weight conditions
$i[\mathbf{K}_{\mu\bar\mu},\mathbb{O}]=0$ for various operators considered in the main text.

\section{}\label{App:UR}

In this Appendix, we collect some useful representations for the coefficient functions
$\Psi_{Nq}$. Let $\vev{*,*}_{j}$  be a $n$-particle $sl(2)$ invariant scalar product
defined as
\begin{align}\label{}
\vev{\Psi,\Phi}_{j}=\prod_{k=1}^n \int_{|z_k|<1}
d^2z_k \,\mu_k(z_k)\, 
\overline{\Psi(z)}\,{\Phi(z)}\,,
\end{align}
where $z_k$ are complex variables, $j=\{j_1,\ldots,j_n\}$ is  a
multi-index and the weight function $\mu_k(z_k)$ is given by the following expression
\begin{align}
\mu_k(z_k)=\dfrac{2j_k-1}{\pi}(1-|z_k|^2)^{2j_k-2}\,.
\end{align}
The generator $S_0=\sum_{i} S_0^{(i)}$ is self-adjoint with respect to this scalar product, $S_0^\dagger=S_0$
and $S_\pm^\dagger=-S_\mp$.
The unit operator (or the reproducing kernel) on this space takes the form
\begin{align}\label{}
\mathbb{K}_j(z,w)=\prod_{k=1}^n(1-z_k\bar w_k)^{-2j_k}\,.
\end{align}
One can easily check that
\begin{align}\label{}
\Phi(z)=\vev{\overline{\mathbb{K}_j(z,w)}|\Phi(w)}_j\,.
\end{align}
We define the $sl(2)$  Fourier transform of the function
$\Psi_{N}(z_1,\ldots,z_n)$ as follows
\begin{align}\label{}
P_{N}(u_1,\ldots,u_n)=\vev{e^{\sum_{k=1}^n z_n u_n}|\Psi_{N}(z)}_j\,.
\end{align}
Notice that
\begin{align}\label{}
\overline{\Psi_N(\partial_{u_1},\partial_{u_n})}P_{N}(u_1,\ldots,u_n)\Big|_{u_i=0}=
||\Psi_N||^2_j\,.
\end{align}
For an arbitrary polynomial $\Psi_N$, one gets the following relation between the polynomials $P_N$ and
$\Psi_N$~\cite{Derkachov:1997qv}
\begin{align*}\label{}
\Psi_N(z)=
\prod_{k=1}^n\frac1{\Gamma(2j_k)}\int_0^\infty du_k{ u_k^{2j_k-1}} e^{-u_k} P_N(u_1 z_1,\ldots,u_n z_n).
\end{align*}
Another useful relation holds for the shift invariant polynomials $\Psi_N$,
$S_- \Psi_N=0$. (Note that the coefficient functions $\Psi_{Nq}$ of conformal
operators entering the expansion of nonlocal operators are always shift invariant.)
This new relation reads
\begin{align}\label{A1}
S_+^k\Psi_{N}(z)&=
\frac{k!}{(N+k)!}\prod_{m=1}^n\frac1{\Gamma(2j_m)}\int_0^\infty du_m u^{2j_m-1}e^{-u_m}
\notag\\
&\quad\times\left(\sum\limits_{p=1}^n u_p z_p \right)^{N+k} \, P_N(u_1,\ldots,u_n)\,.
\end{align}
Taking into account that $P_N(u)$ is a polynomial of the degree $N$ and
separating the so-called infinite volume integration,
$\Lambda=\sum u_i$, we arrive at
\begin{align}\label{A3}
a_{Nk}(j)
S_+^k\Psi_{N}(z)&=
\frac{\chi_{N}(j)}{(N+k)!}\int \mathcal{D}_n x\prod_{m=1}^n x_m^{2j_m-1}
\notag\\[2mm]
&\quad\times\left(\sum\limits_{p=1}^n x_p z_p \right)^{N+k}  P_N(x).
\end{align}
The integral goes on over $N-$dimensional simplex, i.e.
\begin{align}\label{}
\mathcal{D}_n x=\prod_{m=1}^n dx_m\delta\left(1-\sum_{i=1}^nx_i\right)\,,
\end{align}
and the coefficients $a_{Nk}(j)$ and $\chi_{N}(j)$ read
\begin{align*}\label{}
a_{Nk}(j)&=\frac{\Gamma(2N+2J)}{k!\Gamma(2N+2J+k)}\,,
\notag\\
\chi_{N}(j)&=\frac{\Gamma(2N+2J)}{\Gamma(2j_1)\ldots\Gamma(2j_n)},
\end{align*}
where $J=\sum_{i=1}^n j_n$ is the total conformal spin. The representation~(\ref{A3})
allows us to
resum a series over $k$ which appears in the expansion of nonlocal operators. Namely, we have
\begin{multline}\label{A4}
\sum_{k=0}^\infty (-i(pn))^{N+k}\,a_{Nk}(j)S_+^k\Psi_{N}(z)=\\
=\chi_N(j)\int\mathcal{D}_n x
\prod_{m=1}^n x_m^{2j_m-1}\,
e^{-i(pn)\sum_i x_i z_i}
 P_N(x),
\end{multline}
where we take into account that an integral of the polynomial $P_N$ with  any polynomial
of less degree vanishes.

Finally, let us note that there exists the following relation between the $sl(2)$ Fourier
transforms
\begin{align}\label{SSrel}
\vev{e^{u_k z_k}|\Psi}_{\frac12,j_2,j_3\ldots}=\partial_{u_1} u_1\vev{e^{u_k z_k}|\Psi}_{1,j_2,j_3\ldots}
\end{align}
which can easily be checked by expanding the exponent in a power series and taking into
account that
\begin{align}
||z^n||^2_j=\frac{\Gamma(2j)n!}{\Gamma(n+2j)}\,.
\end{align}
Eq.~(\ref{SSrel})  allows one to express polynomials $\mathcal{P}_{Nq}$ entering WW part
of higher twist DAs in terms of the twist$-3$ nucleon functions
$P_{Nq}$.

\section{}\label{App:list}

In this Appendix we collect expressions for the derivatives of the twist$-5$ operators $\mathbb{O}^{t=5,(a)}_{Nq}$
that were used in
sect.~\ref{PF5} in order to derive the equations on  the functions $\Upsilon_{Nq}$.
One easily finds
\begin{align}\label{}
(\bar\lambda\partial_{\bar\mu})\mathbb{O}^{t=5,(1)}_{Nq}&=\mathbb{O}^{t=4,(1)}_{Nq}\,,
\notag\\
(\lambda\partial_{\mu})\mathbb{O}^{t=5,(1)}_{Nq}&=\widetilde{\mathbb{O}}^{t=4,(1)}_{Nq}\,.
\end{align}
For $a=2,\ldots,4$, we obtain
\begin{align}\label{}
(\bar\lambda\partial_{\bar\mu})\mathbb{O}^{t=5,(a)}_{N+1q}&=A^{a}_N\,\mathbb{O}^{t=4,(2)}_{N+1q}
+
B^a_N\,\partial_+\mathbb{O}^{t=4,(1)}_{Nq}\,,
\notag\\
(\lambda\partial_{\mu})\mathbb{O}^{t=5,(a)}_{N+1q}&=\widetilde A^{a}_N\,\widetilde{\mathbb{O}}^{t=4,(2)}_{N+1q}
+
\widetilde B^a_N\,\partial_+\widetilde{\mathbb{O}}^{t=4,(1)}_{Nq}\,,
\end{align}
where
\begin{align}\label{ABAB}
{A}_N&=\Big\{1,\ (2N+5)^{-1},\ 4(N+3)^2\Big\},
\notag\\[1mm]
{B}_N&=\frac{N+2}{2N+5}\Big\{0,\ ((2N+5)(N+3))^{-1},\ 2\Big\},
\notag\\[1mm]
\widetilde{A}_N&=\left\{ \frac{N+4}{(N+2)(2N+5)},\ 1,\ 4(N+3)(N+4)\right\},
\notag\\[1mm]
\widetilde{B}_N&=\frac{N+1}{2N+5}\Big\{((2N+5)(N+2))^{-1},\ 0,\  2\Big\}.
\end{align}
Finally for the last operator we derive
\begin{align}\label{}
&(\bar\lambda\partial_{\bar\mu})\mathbb{O}^{t=5,(5)}_{N+2,q}=C_N
\partial_+\mathbb{O}^{t=4,(2)}_{N+1q}\,,
\notag\\
&(\lambda\partial_{\mu})\mathbb{O}^{t=5,(5)}_{N+2,q}=C_N
\partial_+\widetilde{\mathbb{O}}^{t=4,(2)}_{N+1q}\,,
\end{align}
where
\begin{align}\label{C_N}
C_N=\frac{4(N+3)(N+4)(2N+5)}{2N+7}
\end{align}



\begin{thebibliography}{99}
\bibitem{Chernyak:1977as}
  V.~L.~Chernyak and A.~R.~Zhitnitsky,
  JETP Lett.\  {\bf 25}, 510 (1977)
  [Pisma Zh.\ Eksp.\ Teor.\ Fiz.\  {\bf 25}, 544 (1977)].

\bibitem{Chernyak:1977fk}
  V.~L.~Chernyak, A.~R.~Zhitnitsky and V.~G.~Serbo,
  JETP Lett.\  {\bf 26}, 594 (1977)
  [Pisma Zh.\ Eksp.\ Teor.\ Fiz.\  {\bf 26}, 760 (1977)].

\bibitem{Efremov:1979qk}
  A.~V.~Efremov and A.~V.~Radyushkin,
  Phys.\ Lett.\ B {\bf 94}, 245 (1980).

\bibitem{Efremov:1978rn}
  A.~V.~Efremov and A.~V.~Radyushkin,
  Theor.\ Math.\ Phys.\  {\bf 42}, 97 (1980)
  [Teor.\ Mat.\ Fiz.\  {\bf 42}, 147 (1980)].

\bibitem{Lepage:1979zb}
  G.~P.~Lepage and S.~J.~Brodsky,
  Phys.\ Lett.\ B {\bf 87}, 359 (1979).

\bibitem{Duncan:1979hi}
  A.~Duncan and A.~H.~Mueller,
  Phys.\ Rev.\ D {\bf 21}, 1636 (1980).

\bibitem{Duncan:1979ny}
  A.~Duncan and A.~H.~Mueller,
  Phys.\ Lett.\ B {\bf 90}, 159 (1980).

\bibitem{Milshtein:1981cy}
  A.~I.~Milshtein and V.~S.~Fadin,
  Yad.\ Fiz.\  {\bf 33}, 1391 (1981).

\bibitem{Milshtein:1982js}
  A.~I.~Milshtein and V.~S.~Fadin,
  Yad.\ Fiz.\  {\bf 35}, 1603 (1982).

\bibitem{Kivel:2010ns}
  N.~Kivel and M.~Vanderhaeghen,
  Phys.\ Rev.\ D {\bf 83}, 093005 (2011).


\bibitem{Balitsky:1986st}
  I.~I.~Balitsky, V.~M.~Braun and A.~V.~Kolesnichenko,
  Sov.\ J.\ Nucl.\ Phys.\  {\bf 44}, 1028 (1986)
  [Yad.\ Fiz.\  {\bf 44}, 1582 (1986)].

   \bibitem{Balitsky:1989ry}
  I.~I.~Balitsky, V.~M.~Braun and A.~V.~Kolesnichenko,
  Nucl.\ Phys.\ B {\bf 312}, 509 (1989).

 \bibitem{Chernyak:1990ag}
  V.~L.~Chernyak and I.~R.~Zhitnitsky,
  Nucl.\ Phys.\ B {\bf 345}, 137 (1990).

\bibitem{Braun:2001tj}
  V.~M.~Braun, A.~Lenz, N.~Mahnke and E.~Stein,
  Phys.\ Rev.\  D {\bf 65} (2002) 074011.

\bibitem{Braun:2006hz}
  V.~M.~Braun, A.~Lenz and M.~Wittmann,
  Phys.\ Rev.\ D {\bf 73}, 094019 (2006).

\bibitem{Lenz:2003tq}
  A.~Lenz, M.~Wittmann and E.~Stein,
  Phys.\ Lett.\ B {\bf 581}, 199 (2004).

\bibitem{Aliev:2008cs}
  T.~M.~Aliev, K.~Azizi, A.~Ozpineci and M.~Savci,
  Phys.\ Rev.\ D {\bf 77}, 114014 (2008).

\bibitem{PassekKumericki:2008sj}
  K.~Passek-Kumericki and G.~Peters,
  Phys.\ Rev.\ D {\bf 78}, 033009 (2008).


\bibitem{Anikin:2013aka}
  I.~V.~Anikin, V.~M.~Braun and N.~Offen,
  arXiv:1310.1375 [hep-ph].

\bibitem{Ball:1998ff}
  P.~Ball and V.~M.~Braun,
  Nucl.\ Phys.\ B {\bf 543}, 201 (1999).

\bibitem{Belitsky:2000vx}
  A.~V.~Belitsky and D.~Mueller,
  Nucl.\ Phys.\ B {\bf 589}, 611 (2000).

\bibitem{Radyushkin:2000jy}
  A.~V.~Radyushkin and C.~Weiss,
  Phys.\ Lett.\ B {\bf 493}, 332 (2000).

\bibitem{Kivel:2000rb}
  N.~Kivel, M.~V.~Polyakov, A.~Schafer and O.~V.~Teryaev,
  Phys.\ Lett.\ B {\bf 497}, 73 (2001).

\bibitem{Anikin:2001ge}
  I.~V.~Anikin and O.~V.~Teryaev,
  Phys.\ Lett.\ B {\bf 509}, 95 (2001).

\bibitem{Geyer:2004bx}
  B.~Geyer, D.~Robaschik and J.~Eilers,
  Nucl.\ Phys.\ B {\bf 704}, 279 (2005).

\bibitem{Braun:2000kw}
  V.~Braun, R.~J.~Fries, N.~Mahnke and E.~Stein,
  Nucl.\ Phys.\ B {\bf 589}, 381 (2000)
  [Erratum-ibid.\ B {\bf 607}, 433 (2001)]


\bibitem{Braun:2011aw}
  V.~M.~Braun, T.~Lautenschlager, A.~N.~Manashov and B.~Pirnay,
  Phys.\ Rev.\ D {\bf 83}, 094023 (2011).


\bibitem{Braun:2008ia}
  V.~M.~Braun, A.~N.~Manashov and J.~Rohrwild,
  Nucl.\ Phys.\  B {\bf 807}, 89 (2009).

\bibitem{Braun:2009vc}
  V.~M.~Braun, A.~N.~Manashov and J.~Rohrwild,
  Nucl.\ Phys.\ B {\bf 826}, 235 (2010).


\bibitem{BDKM}
  V.~M.~Braun, S.~E.~Derkachov, G.~P.~Korchemsky and A.~N.~Manashov,
  Nucl.\ Phys.\  B {\bf 553} (1999) 355.

\bibitem{Ball:1998sk}
  P.~Ball, V.~M.~Braun, Y.~Koike and K.~Tanaka,
  Nucl.\ Phys.\ B {\bf 529}, 323 (1998).

\bibitem{Derkachov:1997qv}
  S.~E.~Derkachov, S.~K.~Kehrein and A.~N.~Manashov,
  Nucl.\ Phys.\ B {\bf 493}, 660 (1997).

\end{thebibliography}
\end{document}